\documentclass[reprint,nofootinbib,aps,superscriptaddress,amsmath,amssymb,onecolumn,11pt]{revtex4-2}

\usepackage{graphicx}
\usepackage{dcolumn}
\usepackage{bm}
\usepackage{breqn}
\usepackage{color}

\usepackage[colorlinks=true,linkcolor=red,citecolor=blue]{hyperref}
\DeclareRobustCommand{\rev}[1]{{\color{black}#1}}

\begin{document}

\title{Electric Penrose process in spherically symmetric regular black holes with and without a cosmological constant}

\author{Haowei Chen}
\email{chenhaowei@zjut.edu.cn}
\author{Hengyu Xu}
\email{xuhengyu0501@outlook.com}
\author{Yizhi Zhan}
\email{niannian\_12138@163.com}
\author{Shao-Jun Zhang}
\email{sjzhang@zjut.edu.cn (corresponding author)}
\affiliation{Institute for Theoretical Physics and Cosmology$,$ Zhejiang University of Technology$,$ Hangzhou 310032$,$ China}
\affiliation{United Center for Gravitational Wave Physics$,$ Zhejiang University of Technology$,$ Hangzhou 310032$,$ China}
\date{\today}

\begin{abstract}
We investigate the electric Penrose process in Ay\'{o}n-Beato-Garc\'{i}a (ABG) black holes, both in the presence and absence of a cosmological constant, presenting, to the best of our knowledge, the first such analysis within the context of regular black holes. Our study systematically examines the effects of black hole charge and the cosmological constant on the formation of negative-energy states and the efficiency of energy extraction. Compared to Reissner-Nordstr\"{o}m (RN) black holes, ABG black holes exhibit a significantly larger negative-energy region, enabling the electric Penrose process to operate at larger distances from the event horizon and achieve higher energy extraction efficiency. This enhancement is particularly pronounced near the event horizon, where the performance gap widens with increasing black hole charge. Notably, even for astrophysically realistic values of charge and cosmological constant that approach vanishingly small values, distinct differences persist, yielding a maximum efficiency ratio of approximately $23/8$. These results suggest that, in realistic astrophysical scenarios, ABG black holes can accelerate charged particles more efficiently and serve as more powerful engines for energy extraction than their RN counterparts.
\end{abstract}


\maketitle
  
\section{Introduction}

In recent years, alongside remarkable advancements in astronomical observations, black hole (BH) physics has emerged as a frontier focus in contemporary physics research. As the most compact astrophysical objects predicted by general relativity (GR), BHs are widely recognized as the ``engines" driving a range of high-energy astrophysical phenomena, including relativistic jets in active galactic nuclei and gamma-ray bursts. The immense energy unleashed by these phenomena is thought to originate either from the intrinsic energy of BHs themselves or from the gravitational potential energy released as matter accretes onto them.

According to BH thermodynamics, a Kerr BH contains an enormous amount of extractable energy. For the extremal case, this energy can reach up to $0.29 M c^2$ \cite{Christodoulou:1970wf, Bekenstein:1973mi}, where $M$ denotes the BH's mass and $c$ is the speed of light in a vacuum. How to extract such vast energy has long been a key theoretical focus in BH physics. The first mechanism for BH energy extraction was proposed by Penrose and is known as the Penrose process \cite{Penrose:1971uk}. This process involves a particle splitting into two within the ergoregion of a rotating BH. One particle, carrying negative energy, falls into the BH, while the other escapes with excess energy, thereby extracting energy from the BH. Although theoretically compelling, the Penrose process requires the relative velocity of the two split particles to exceed half the speed of light \cite{Bardeen:1972fi, Wald:1974kya}. Such a scenario is exceedingly rare in realistic astrophysical processes. Moreover, its energy extraction efficiency is low, especially for BHs with low spin, which means it lacks practical feasibility. Drawing inspiration from the Penrose process, researchers have subsequently proposed various modified or alternative mechanisms. These include the magnetic Penrose process \cite{dhurandharEnergyextractionProcessesKerr1984,dhurandharEnergyextractionProcessesKerr1984a,1985ApJ...290...12W,parthasarathyHighEfficiencyPenrose1986,Bhat:1985hpc,Wagh:1989zqa}, the collisional Penrose process \cite{piran1975high}, superradiant scattering \cite{Teukolsky:1974yv}, the Blandford-Znajek mechanism \cite{Blandford:1977ds}, the magnetohydrodynamic Penrose process \cite{1990ApJ...363..206T}, the Ba\~{n}ados-Silk-West mechanism \cite{Banados:2009pr}, and the Comisso-Asenjo mechanism \cite{Comisso:2020ykg}. Some of these mechanisms overcome the limitations of the Penrose process and stand as strong candidate explanations for high-energy astrophysical phenomena.

The Penrose process and its various variants typically rely on the existence of a BH's ergoregion, where particles can occupy negative-energy orbits. For non-rotating BHs such as Schwarzschild BHs, the Penrose process cannot occur due to the absence of an ergoregion. In realistic astrophysical environments, BHs are generally considered incapable of carrying electric charge because of rapid discharge processes. Some studies, however, suggest that in certain astrophysical scenarios, BHs may carry a small amount of electric charge \cite{Ruffini:1975ne, 1978ApJ...220..743B, Weingartner:2006rq, Wald:1974np, Levin:2018mzg, Tursunov:2019mox, Tursunov:2016dss, Zajacek:2018ycb, Zajacek:2019kla}. Although the backreaction of the charge on the spacetime geometry is negligible, it alters the motion of charged particles significantly. For a Reissner-Nordstr\"{o}m (RN) BH, it still lacks an ergoregion. The electric field, though, creates a region outside the event horizon analogous to an ergoregion for charged particles. This can be termed a generalized ergoregion or negative-energy region. Within this generalized ergoregion, charged particles can occupy negative-energy orbits. This enables an energy extraction process similar to the Penrose process, which is referred to as the electric Penrose process \cite{Tursunov:2021jjf, Zaslavskii:2024zgh}. In this process, the split charged particles can be accelerated to extremely high energies. Research has found that even a minor charge on the BH can achieve ultra-high energy extraction efficiency. This may help account for ultra-high-energy cosmic ray phenomena \cite{Tursunov:2021jjf}. This study has been extended to include cases involving a cosmological constant \cite{Baez:2024lhn}.

On the other hand, classical GR predicts spacetime singularities or geodesic incompleteness \cite{Penrose:1964wq, Hawking:1973uf, Senovilla:1998oua}. The existence of singularities indicates the failure of classical theories in this context. It is widely believed that quantum gravitational or matter effects may resolve spacetime singularities. In the absence of a complete theory of quantum gravity, nonlinear electrodynamics provides a classical effective framework that can phenomenologically mimic certain features expected from quantum vacuum polarization in strong-field regimes \cite{Heisenberg:1936nmg, Born:1934gh, Polchinski:1998rq, Polchinski:1998rr}. In recent years, researchers have discovered that by considering Einstein-nonlinear electromagnetic field theories, singularity-free BH solutions can be obtained \cite{Ayon-Beato:1998hmi, Ayon-Beato:1999kuh, Ayon-Beato:2000mjt}. Based on this approach, researchers have constructed numerous regular BH solutions. These include the Bardeen BH \cite{1968qtr..conf...87B, Ayon-Beato:2000mjt} and Ay\'{o}n-Beato-Garc\'{i}a (ABG) BH \cite{Ayon-Beato:1998hmi, Ayon-Beato:1999kuh}, among others. The properties of these regular BHs and their applications in astrophysics have also been extensively studied. For reviews on this topic, see \cite{Lan:2023cvz,Torres:2022twv,Bronnikov:2022ofk,Cai:2020kue,Cai:2021ele,Yang:2022uze,Meng:2022oxg}. Research results show that regular BHs exhibit unique characteristics distinct from ordinary BHs. For example, RN BHs do not exhibit superradiant instability under charged scalar field perturbations \cite{Hod:2012wmy,Hod:2013nn}. Regular charged BHs, however, possess this feature \cite{Hod:2024aen,dePaula:2024xnd,Dolan:2024qqr, Zhan:2024gvi}. This reflects the significant influence of electromagnetic field nonlinearity on BH properties.

Building on these studies, the present work presents, to the best of our knowledge, the first systematic investigation of the electric Penrose process within the framework of regular black holes, specifically focusing on the Ay\'{o}n-Beato-Garc\'{i}a (ABG) black hole family. This study therefore extends previous analyses restricted to singular black hole geometries. We further extend our analysis to incorporate the effects of the cosmological constant, considering both asymptotically flat ABG black holes and ABG-de Sitter (ABG-dS) black holes. \rev{The electric Penrose process may be regarded as the particle analogue of charged superradiance \cite{DiMenza:2015raa}. Their common physical origin is the electromagnetic coupling of charged matter to the BH electrostatic potential. In the electric Penrose process, this coupling allows a negative-Killing-energy fragment to enter the horizon, leaving the escaping fragment with more energy than the incident particle. In charged superradiance, it permits a negative energy flux through the horizon in the superradiant regime, thereby amplifying the reflected wave.} Consequently, we anticipate that the electric Penrose process in ABG(-dS) black holes will exhibit distinctive characteristics that differ fundamentally from those in RN(-dS) black holes, potentially offering a novel observational signature to differentiate between singular and regular black hole spacetimes.

The structure of this work is as follows. In Sec. II, we analyze the motion of charged particles in general spherical charged BHs and introduce the effective potential and negative-energy regions. In Sec. III, we derive the general formalism of the electric Penrose process. In Secs. IV and V, we apply this general formalism to ABG BHs and ABG-dS BHs, respectively, and investigate their energy extraction efficiencies. The final section presents a summary and discussion.

Throughout this work, we use the units $c = G = 4 \pi \epsilon_0 = 1$, where $c$, $G$, and $\epsilon_0$ denote the speed of light in vacuum, Newton's gravitational constant, and the vacuum permittivity, respectively.

\section{Motion of charged particles in charged black holes}

We consider charged particles moving in static, spherically symmetric charged BHs. The general form of the metric is 
\begin{align}
    ds^2 = -f(r) dt^2 + \frac{1}{f(r)} dr^2 + r^2 (d\theta^2 + \sin^2\theta d \phi^2),
\end{align}
where $f(r)$ is the metric function. The electromagnetic potential takes the form
\begin{align}
    A_\mu = (A_t(r), 0, 0, 0).
\end{align}
For RN-(dS) BHs, $f(r) = 1- \frac{2 M}{r} + \frac{Q^2}{r^2} - \frac{\Lambda r^2}{3}$ and $A_t (r)= - \frac{Q}{r}$, where $M$ denotes the BH's mass, $Q$ is the BH's charge, and $\Lambda$ is the cosmological constant. This scenario has been discussed in \cite{Tursunov:2021jjf, Baez:2024lhn}. In the present work, we focus on the ABG(-dS) BHs, with their corresponding $f(r)$ and $A_t$ given below.

\subsection{Equations of motion}

The dynamics of a charged particle is governed by the following Lagrangian
\begin{align}
    {\cal L} = \frac{1}{2} m g_{\mu \nu} \dot{x}^\mu \dot{x}^\nu + q A_\mu \dot{x}^\mu, \label{Lagrangian}
\end{align}
where $m$ and $q$ denote the mass and electric charge of the particle, respectively. The dot symbol represents the derivative with respect to the proper time $\tau$, i.e., $\dot{x}^\mu \equiv d x^\mu /d\tau$. Owing to the spacetime symmetry, the Lagrangian (\ref{Lagrangian}) does not explicitly depend on the coordinates $\{t, \phi\}$. Consequently, their conjugate momenta are conserved quantities, expressed as
\begin{align}
    p_t  \equiv \frac{\partial {\cal L}}{\partial \dot{t}} & = m U_t + q A_t \equiv - E, \label{Energy}\\
    p_\phi \equiv \frac{\partial {\cal L}}{\partial \dot{\phi}} &= m U_\phi \equiv L, \label{AngularMomentum}
\end{align}
where $U^\mu \equiv \dot{x}^\mu$ is the 4-velocity of the particle. The constants $E$ and $L$ correspond to the energy and the $z$-component of the angular momentum of the particle, respectively, as observed by observers at infinity.

Without loss of generality, we assume the particle initially moves in the equatorial plane $\theta = \pi/2$. Due to the spacetime symmetry, the particle will remain confined to this plane. Hence, in subsequent calculations, we fix $\theta = \pi/2$ with $\dot{\theta} = 0$. The 4-velocity of the particle may be expressed as $U^\mu = \dot{t} (1, v, 0, \Omega)$, where $v \equiv dr/dt$ and $\Omega \equiv d\phi/dt$ denote the radial and angular velocities measured at infinity, respectively. Using this notation, the renormalization condition for the 4-velocity, $U_\mu U^\mu = -1$, takes the form
\begin{align}
    g_{tt} + g_{rr} v^2 + g_{\phi\phi} \Omega^2 = -\frac{1}{(\dot{t})^2},
\end{align}
from which the angular velocity $\Omega$ can be derived as
\begin{align}
    \Omega = \pm \frac{1}{r} \sqrt{f(r) - f^{-1}(r) v^2 - \frac{1}{(\dot{t})^2}}. \label{AngularVelocity}
\end{align}
This expression reveals that for any physical particle, the permissible range of angular velocity $\Omega$ is bounded by
\begin{align}
    \Omega_- \leq \Omega \leq \Omega_+, \qquad \Omega_\pm = \pm \frac{1}{r} \sqrt{f(r) - f^{-1}(r) v^2},
\end{align}
where $\Omega_\pm$ correspond to photon motion (i.e., satisfying $U_\mu U^\mu =0$).

\subsection{Effective potential and negative-energy regions}

Using Eqs.~(\ref{Energy}) and~(\ref{AngularMomentum}), the renormalization condition of the 4-velocity, $U_\mu U^\mu = -1$, can be written as a quadratic equation in terms of the specific energy $e \equiv E/m$,
\begin{align}
    e^2 - 2 \beta e + \gamma=0,
\end{align}
where 
\begin{align}
    \beta &= - \bar{q} A_t,\\
    \gamma &= \bar{q}^2 A_t^2  - f(r) \left(\frac{\ell^2}{r^2} +1 \right)- \dot{r}^2.
\end{align}
Here, $\bar{q} \equiv q/m$ denotes the charge-to-mass ratio of the particle, and $\ell \equiv L/m$ is the specific angular momentum of the particle as measured from infinity. Solving this quadratic equation yields 
\begin{align}
    e =  -\bar{q} A_t + \sqrt{f(r) \left(\frac{\ell^2}{r^2} +1 \right) + \dot{r}^2}. \label{Energy1}
\end{align}
The positive sign for the square root is chosen to ensure $U^t >0$, guaranteeing that the particle always moves forward in time relative to local static observers (For more details on this point, refer to \cite{Gupta:2021vww, Zhang:2025blr}). Using this expression for the energy, we define the effective potential as
\begin{align}
    V_{\rm eff} (r) \equiv e(\dot{r}=0) = -\bar{q} A_t + \sqrt{f(r) \left(\frac{\ell^2}{r^2} +1\right)},\label{EffectivePotential}
\end{align}
which represents the minimum specific energy that the particle must possess. In other words, a particle with specific energy $e$ can only move in regions where $e \geq V_{\rm eff}(r)$. 

Energy extraction via the Penrose process depends on the existence of a negative-energy region (NER), where $V_{\rm eff} (r)<0$ (making it possible for $e<0$). In the absence of an electromagnetic field ($A_t =0$), $V_{\rm eff} (r)$ (and thus $e$) is always non-negative, meaning no negative-energy region exists, and energy extraction through the Penrose process is therefore impossible. However, when the electromagnetic interactions are included, the first term in Eq.~(\ref{EffectivePotential}) can give rise to negative-energy regions outside the horizon, defined by the condition
\begin{align}
    V_{\rm eff} (r) < 0. \label{NER}
\end{align}
Within such regions, the electric version of the Penrose process can occur. 

\section{Electric Penrose process}

We now consider energy extraction from the BH via the electric Penrose process. For simplicity, we examine a straightforward scenario: a particle (labeled $1$) moving in the equatorial plane splits into two fragments (labeled $2$ and $3$), which also propagate within the equatorial plane. In this process, both 4-momentum and electric charge are conserved, leading to the relations
\begin{align}
 m_1 U^\mu_1 &= m_2 U^\mu_2 + m_3 U^\mu_3, \label{MomentumConservation}\\
 q_1 & = q_2 + q_3. \label{ChargeConservation}
\end{align}
Here, the subscripts $1, 2, 3$ denote the three particles, respectively. The above conservation equations imply the conservation of energy and angular momentum, specifically
\begin{align}
    E_1 &= E_2 + E_3,\label{EnergyConservation}\\
    L_1 &= L_2 + L_3.\label{AngularMomentumConservation}
\end{align}
Energy extraction occurs when particle $2$ acquires negative energy ($E_2<0$) and falls into the BH, while particle $3$ escapes to infinity with excess energy ($E_3 > E_1$) to maintain total energy conservation. 


The efficiency of the energy extraction is defined as
\begin{align}
    \eta \equiv \frac{E_3 - E_1}{E_1}. \label{Efficiency-1}
\end{align}
Using Eqs.~(\ref{MomentumConservation}) and~(\ref{Energy}), the efficiency can be expressed as \cite{Tursunov:2021jjf, Zhang:2025blr}
\begin{align}
    \eta = \kappa -1 + \kappa \frac{\bar{q}_1 A_t}{e_1} - \frac{m_3}{m_1}\frac{\bar{q}_3 A_t}{e_1}, \label{Efficiency-2}
\end{align}
where
\begin{align}
    \kappa = \frac{\Omega_1 - \Omega_2}{\Omega_3 - \Omega_2}.
\end{align}
Following \cite{1986ApJ...307...38P}, we introduce two simplifying assumptions. First, we set $q_1 =0$, such that $q_2 = - q_3 \equiv q$. \rev{This choice represents the ionization or decay of a neutral parent into oppositely charged fragments and defines the reference channel used below.} Second, we adopt the approximation $e_1 \approx 1$, which is valid since the incident particle moves non-relativistically in nearly all realistic scenarios~\cite{1986ApJ...307...38P}. With this approximation, the angular velocity of particle $1$ can be derived as
\begin{align}
    \Omega_1 = \frac{1}{r} \sqrt{f(r) (1-f(r))}.
\end{align}
Here, we have defined the rotation direction of particle 1 as the positive rotational direction. As discussed in \cite{1986ApJ...307...38P, Tursunov:2021jjf}, $\kappa$ (and consequently $\eta$) is maximized when all radial velocities vanish ($v_i = 0$ for $i=1,2,3$) and the angular velocities of the two fragments taking their limiting values ($\Omega_2 = \Omega_-$ and $\Omega_3 = \Omega_+$). Under these conditions, the efficiency simplifies to 
\begin{align}
    \eta= \frac{1}{2} \left[\sqrt{1-f(r)} - 1\right] + \hat{\bar{q}} A_t. \label{Efficiency-3}
\end{align}
Here, $\hat{\bar{q}}=\frac{m_2}{m_1}\bar{q}$ is referred to as the rescaled charge-to-mass ratio of particle 2. Additionally, the mass ratio $m_2/m_1$ is constrained by $m_2 + m_3 \leq m_1$, a condition that arises from $4$-momentum conservation in the local rest frame of particle $1$. 

\begin{table}[!htbp]
\centering
\caption{\rev{Charge and energy notation used in the manuscript.}}
\label{tab:charge-notation}
\begingroup
\setlength{\tabcolsep}{0pt}
\renewcommand{\arraystretch}{1.15}
\begin{tabular}{@{}c@{\hspace{0.035\linewidth}}l@{}}
\hline\hline
\makebox[0.14\linewidth][c]{\rev{Symbol}} & \makebox[0.68\linewidth][l]{\rev{Definition and convention}}\\
\hline
\makebox[0.14\linewidth][c]{\rev{$Q$}} & \parbox[c]{0.68\linewidth}{\raggedright\rev{Electric charge parameter of the BH.}\strut}\\
\makebox[0.14\linewidth][c]{\rev{$q_i$}} & \parbox[c]{0.68\linewidth}{\raggedright\rev{Electric charge of particle $i$; $q_1=0$ and $q_2=-q_3\equiv q$ in the reference channel.}\strut}\\
\makebox[0.14\linewidth][c]{\rev{$\bar{q}_i$}} & \parbox[c]{0.68\linewidth}{\raggedright\rev{Charge-to-mass ratio $\bar{q}_i\equiv q_i/m_i$; $\bar q$ denotes $\bar q_2$.}\strut}\\
\makebox[0.14\linewidth][c]{\rev{$\hat{\bar q}$}} & \parbox[c]{0.68\linewidth}{\raggedright\rev{Rescaled ratio $\hat{\bar q}\equiv(m_2/m_1)\bar q=q_2/m_1$.}\strut}\\
\makebox[0.14\linewidth][c]{\rev{$e_i$}} & \parbox[c]{0.68\linewidth}{\raggedright\rev{Specific energy $e_i\equiv E_i/m_i$.}\strut}\\
\hline\hline
\end{tabular}
\endgroup
\end{table}

From Eq.~(\ref{Efficiency-3}), the efficiency receives contributions from two terms: the first is a purely geometric factor, while the second originates from electromagnetic interactions. Notably, in the absence of electromagnetic interactions ($A_t = 0$), $\eta$ is always negative, indicating that energy extraction cannot occur for neutral spherical BHs. Energy extraction becomes possible only when electromagnetic interactions are included.

\section{Ay\'{o}n-Beato-Garc\'{i}a black hole}

The metric of the Ay\'{o}n-Beato-Garc\'{i}a (ABG) BH is \cite{Ayon-Beato:1998hmi, Ayon-Beato:1999kuh}
\begin{align}
f(r) &= 1- \frac{2 M r^2}{(r^2 + Q^2)^{3/2}} + \frac{Q^2 r^2}{(r^2 + Q^2)^2},\label{MetricABG}\\
A_t(r) &= - \frac{r^5}{2 Q} \left(\frac{3 M }{r^5} + \frac{2 Q^2}{(r^2 + Q^2)^3} - \frac{3 M}{(r^2 + Q^2)^{5/2}}\right). \label{ElectricPotentialABG}
\end{align}
This constitutes an exact, regular solution to an Einstein-nonlinear-electromagnetic field theory. The parameters $M$ and $Q$ correspond to the mass and the electric charge of the spacetime, respectively. The event horizon, denoted $r=r_h$, is given by the largest root of the equation $f(r)=0$. A horizon exists only when $|Q| \leq Q_{\rm ext} \approx 0.6341 M$, for which the metric describes a regular BH. Asymptotically, the ABG BH behaves like the RN BH. \rev{However, in the small-$Q$ limit, its electric potential takes the form $A_t(r)=-Q/r-15MQ/(4r^2)+{\cal O}(Q^3)$, which differs from the RN potential $A_t^{\rm RN}(r)=-Q/r$.} In what follows, we set $M=1$, such that all physical quantities are measured in units of $M$. Additionally, we take $Q>0$ without loss of generality.

\begin{figure}[!htbp]
    \centering
    \includegraphics[width=0.7\linewidth]{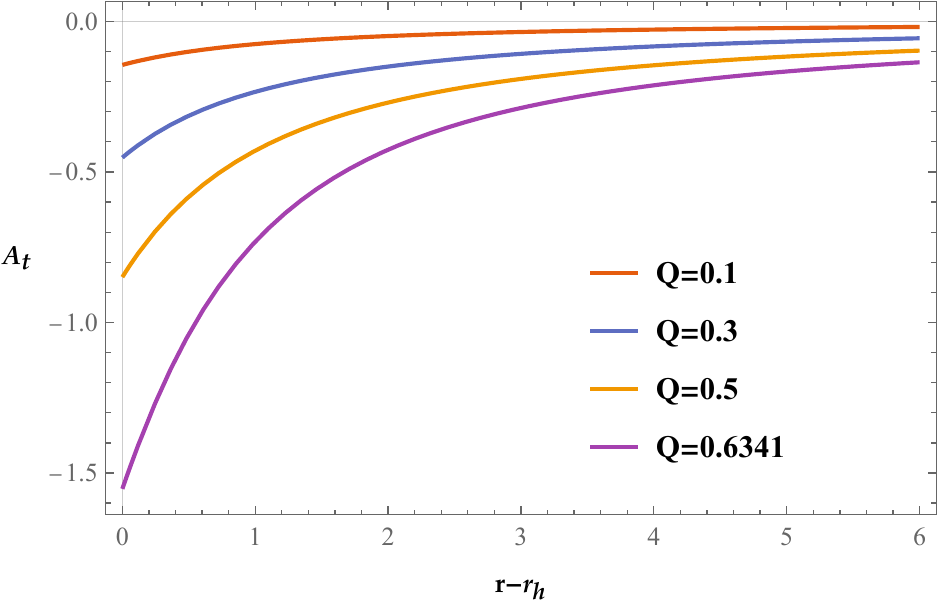}
    \caption{The profile of the electric potential $A_t$ for various $Q$ in the ABG BH. The monotonic negative potential outside the horizon provides the electromagnetic source of negative-energy states for oppositely charged particles.}
    \label{At-r}
\end{figure}

Based on Eq.~(\ref{ElectricPotentialABG}), the profiles of $A_t(r)$ for different values of $Q$ are illustrated in Fig. \ref{At-r}. Within the range $0 < Q \leq Q_{\rm ext} \approx 0.6341 M$, the potential $A_t(r)$ is a monotonically increasing function of $r$. Furthermore, it remains negative outside the horizon and approaches zero as $r \rightarrow \infty$. For a fixed $r$, $A_t$ decreases monotonically with increasing $Q$.

As a consequence, Eq.~(\ref{EffectivePotential}) implies that the existence of a negative-energy region outside the horizon requires the particle to carry a negative charge $q$. In other words, the particle must have an opposite charge sign to that of the BH. This is physically intuitive: the attractive potential between opposite charges is negative, whereas the repulsive potential between like charges is positive. Additionally, when $q<0$, a negative-energy region can always exist in the vicinity of the event horizon. \rev{For $\bar q<0$ and $A_t<0$, the term $-\bar q A_t$ in Eq.~(\ref{EffectivePotential}) is negative. Over the parameter range considered here, the nonlinear-electrodynamic correction makes the ABG potential more negative than the RN potential. It therefore lowers $V_{\rm eff}$ and moves the zero-energy surface outward. This is the direct physical origin of the larger ABG negative-energy region.}

\subsection{Negative-energy regions}

We first investigate the existence of the negative-energy region (NER), as defined by Eq.~(\ref{NER}). From Eqs.~(\ref{EffectivePotential}),~(\ref{MetricABG}), and~(\ref{ElectricPotentialABG}), it is evident that $V_{\rm eff} (r)$ depends on three parameters: $\{Q, \bar{q}, \ell\}$. Only when $\bar{q} <0$ can $V_{\rm eff} (r)$ become negative in some region outside the horizon.

\begin{figure}[!htbp]
	\centering
 	\includegraphics[width=0.46\textwidth]{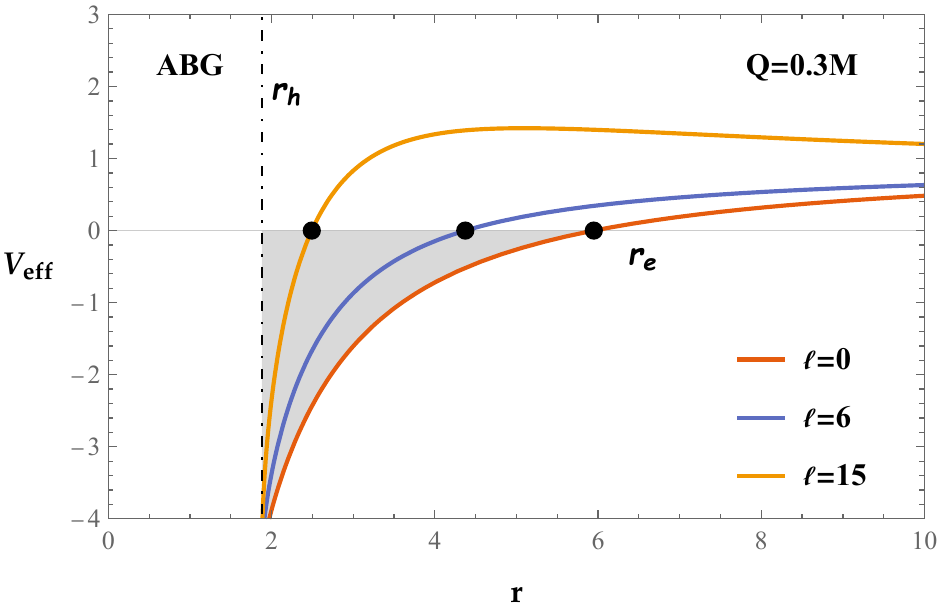}\quad
          \includegraphics[width=0.46\textwidth]{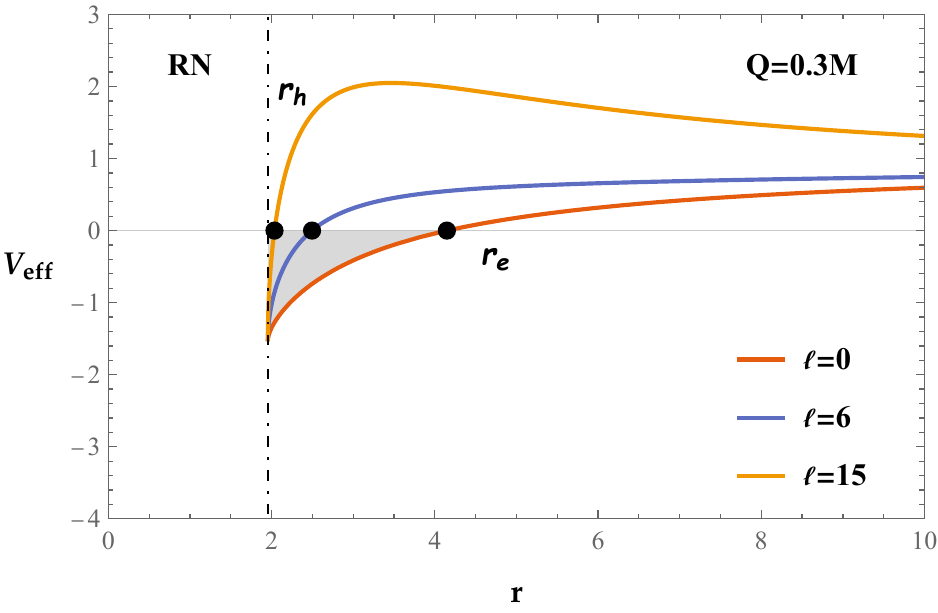}
          
\caption{Effective potential $V_{\rm eff} (r)$ for the ABG (left panel) and RN BHs (right panel) for various $\ell$. The shaded gray regions indicate the negative-energy regions (NER). $r_h$ (dotted-dashed lines) denotes the event horizon, and $r_e$s (black dots) represents the radii of zero-energy surfaces ($V_{\rm eff} (r_e) = 0$). We set $Q=0.3$ and $\bar{q} = -10$. The wider shaded region in the ABG case shows that negative-energy states can extend farther from the horizon.}
\label{VeffNER}
\end{figure}

The left panel of Fig. \ref{VeffNER} illustrates the profile of $V_{\rm eff} (r)$ for representative values of these parameters. $V_{\rm eff} (r)$ exhibits similar behavior for other parameter values. For convenient comparison, the right panel shows the corresponding RN case. The figure clearly demonstrates the presence of a NER near the horizon. As $\ell$ increases, the NER shrinks monotonically, consistent with Eq.~(\ref{EffectivePotential}). For fixed $Q (\leq Q_{\rm ext}), \bar{q}$ and $\ell$, the NER in the ABG case is larger than in the RN case, suggesting that energy extraction may occur at greater distances from the BH. Additionally, since $V_{\rm eff} (r)$ contains only the factor $\ell^2$, motion in clockwise and counterclockwise directions is equivalent, meaning the direction of angular momentum does not affect the effective potential.

\begin{figure}[htbp]
    \centering
    \includegraphics[width=0.7\linewidth]{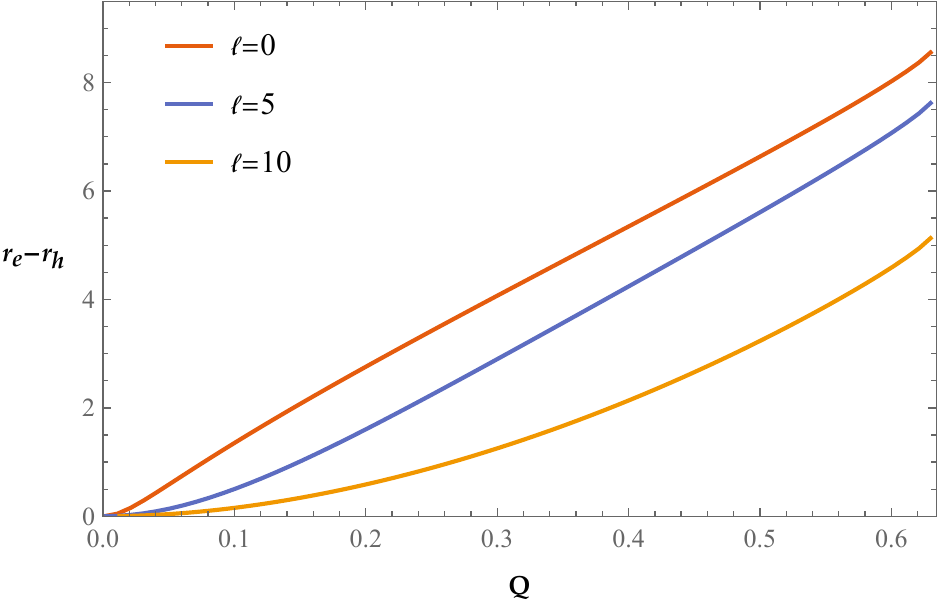}
    \caption{The distance $r_e-r_h$ from the event horizon to the zero-energy surface as a function of the BH charge $Q$ for various $\ell$. We set $\bar{q}=-10$. Its growth with $Q$ shows that the NER widens as the BH charge increases.}
    \label{re-Q}
\end{figure}

Fig. \ref{re-Q} plots the radius of the zero-energy surface $r_e$ against the BH charge $Q$. The figure shows that $r_e-r_h$ increases with $Q$, indicating that the NER expands as the BH charge grows.

The influence of $\bar{q}$ on the NER is evident from Eq.~(\ref{EffectivePotential}): for positive $Q$, as $\bar{q}$ gets more negative, the NER enlarges.

\subsection{Energy extraction efficiency}

\begin{figure}[!htbp]
    \centering
    \includegraphics[width=0.7\linewidth]{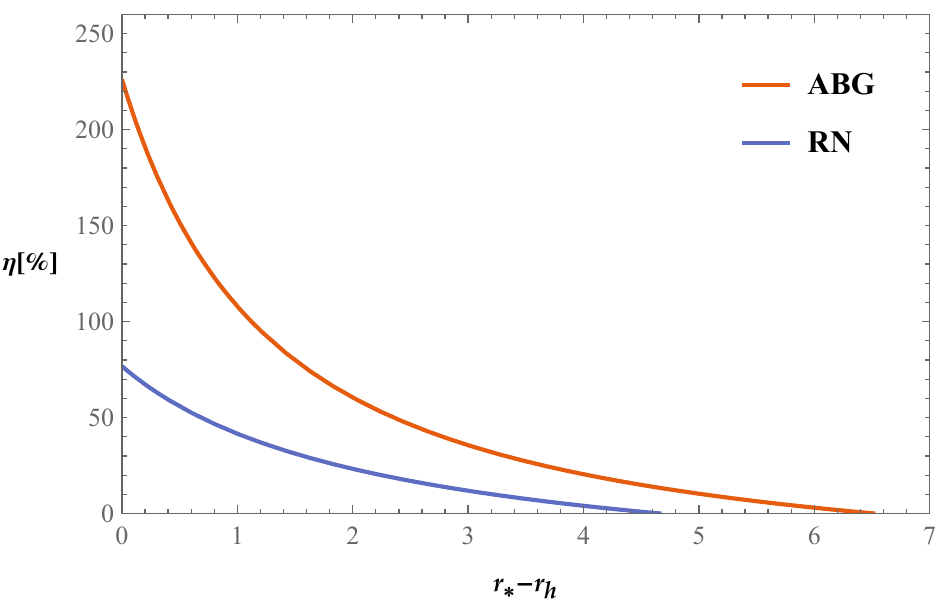}
    \caption{Energy extraction efficiency $\eta$ as a function of the splitting point $r_\ast$. We set $Q=0.3M$ and $\hat{\bar{q}}=-5$. The ABG--RN difference is largest near the event horizon.}
    \label{etarABGRN}
\end{figure}

From Eq.~(\ref{Efficiency-3}), the energy extraction efficiency $\eta$ depends on three parameters: $\{Q, \hat{\bar{q}}\}$ and the splitting point $r_\ast$. The analytical expression clearly reveals the influences of these parameters on $\eta$: $\eta$ increases with the growth of $|Q|$ or $|\hat{\bar{q}}|$, but decreases as $r_\ast$ increases. \rev{The linear growth with $|\hat{\bar q}|$ is formal. In any specified breakup channel, $m_2+m_3\leq m_1$ gives $|\hat{\bar q}|=|q_2|/m_1\leq |q_2|/(m_2+m_3)\leq|\bar q_2|$, so the attainable value is finite and channel dependent. Since elementary particles have enormous charge-to-mass ratios in geometrized units, this finite bound can nevertheless allow $\eta\gg1$.} As shown in Fig. \ref{etarABGRN}, for the selected parameter values, $\eta$ can exceed $100\%$ when the splitting point $r_\ast$ is close to the horizon. For fixed parameters, the efficiency $\eta$ of the ABG BH is consistently higher than that of the RN BH, particularly when $r_\ast$ is near the horizon.

\rev{For macroscopic fragments, electrostatic self-energy and charge stability impose additional model-dependent limits. Moreover, $\Delta E_{\rm max}\simeq |q_2A_t(r_h)|$, so a large dimensionless efficiency does not imply unlimited absolute energy. Pair production, plasma screening, radiation, and BH neutralization further restrict realizable events} \cite{Ruffini:1975ne,Tursunov:2021jjf,Zajacek:2018ycb,Zajacek:2019kla}. \rev{The efficiencies shown below are therefore idealized single-event benchmarks, not predictions of luminosity or event rate.}

\begin{figure}[!htbp]
    \centering
    \includegraphics[width=0.7\linewidth]{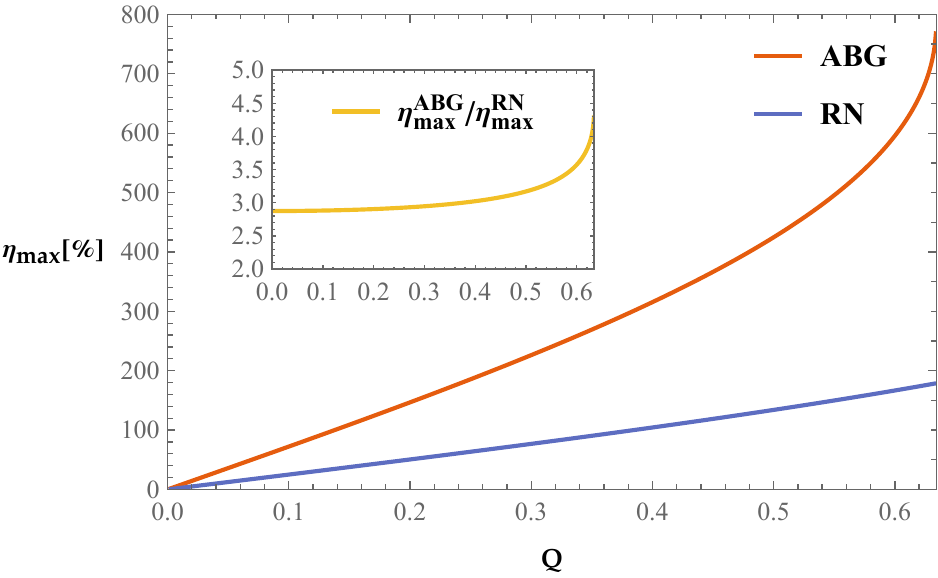}
    \caption{The maximum energy extraction efficiency $\eta_{\rm max}$ as a function of the BH charge $Q$, with the splitting point $r_\ast = r_h$. We set $\hat{\bar{q}}=-5$. The separation between the ABG and RN curves quantifies the stronger extraction capability of the regular geometry.}
    \label{etaQ}
\end{figure}

For fixed $Q$ and $\hat{\bar{q}}$, $\eta$ attains its maximum value $\eta_{\rm max}$ as $r_\ast \rightarrow r_h$. From Eq.~(\ref{Efficiency-3}), we obtain
\begin{align}
    \eta_{\rm max} = \hat{\bar{q}} A_t\big|_{r=r_h}. \label{etaMax}
\end{align}
\rev{The horizon-limit comparison is not restricted to the reference channel. For the same maximum-efficiency kinematics and a noncritical incident particle satisfying $e_1+\bar q_1A_t(r_h)>0$, $\kappa\to1$ as $r_\ast\to r_h$, so Eq.~(\ref{Efficiency-2}) gives $\eta_h=q_2A_t(r_h)/E_1$. Hence, for fixed $E_1$ and $q_2$, the ABG--RN ratio is determined by the corresponding horizon potentials, whereas finite-radius efficiencies remain channel dependent.}
Fig. \ref{etaQ} plots the maximum efficiency $\eta_{\rm max}$ as a function of $Q$. It is evident that $\eta_{\rm max}$ for the ABG BH is consistently higher than that for the RN BH, with the discrepancy increasing as $Q$ grows. When $Q \rightarrow Q_{\rm ext}$, the ratio $\eta_{\rm max}^{\rm ABG} / \eta_{\rm max}^{\rm RN} \sim 4.3$. Interestingly, for small values of $Q$ (even extremely tiny ones), $\eta_{\rm max}$ exhibits a nearly linear dependence on $Q$ in both the ABG and RN cases. \begin{samepage}
\rev{For $Q/M\ll1$,}
\begingroup
\begin{align}
r_h^{\rm ABG}&=2M+{\cal O}(Q^2/M),&
A_t^{\rm ABG}(r_h^{\rm ABG})&=-\frac{23}{16}\frac{Q}{M}
   +{\cal O}(Q^3/M^3),\nonumber\\
r_h^{\rm RN}&=2M+{\cal O}(Q^2/M),&
A_t^{\rm RN}(r_h^{\rm RN})&=-\frac{1}{2}\frac{Q}{M}
   +{\cal O}(Q^3/M^3). \label{SmallQHorizonPotential}
\end{align}
\endgroup
\end{samepage}
\rev{Equation~(\ref{etaMax}) directly gives, for the same $\hat{\bar{q}}$, the ratio of the maximum efficiencies as the ratio of the corresponding horizon potentials:}
\begingroup
\begin{align}
\frac{\eta_{\rm max}^{\rm ABG}}{\eta_{\rm max}^{\rm RN}}
=\frac{\hat{\bar{q}}A_t^{\rm ABG}(r_h^{\rm ABG})}
{\hat{\bar{q}}A_t^{\rm RN}(r_h^{\rm RN})}
=\frac{23}{8}+{\cal O}(Q^2/M^2). \label{SmallQRatio}
\end{align}
\endgroup
\rev{A small positive cosmological constant shifts the common neutral horizon only by ${\cal O}(\Lambda M^3)$, giving corrections of order $\Lambda M^2$ to this ratio.}

\section{Ay\'{o}n-Beato-Garc\'{i}a-de Sitter black hole}

\begin{figure}[!htbp]
    \centering
    \includegraphics[width=0.7\linewidth]{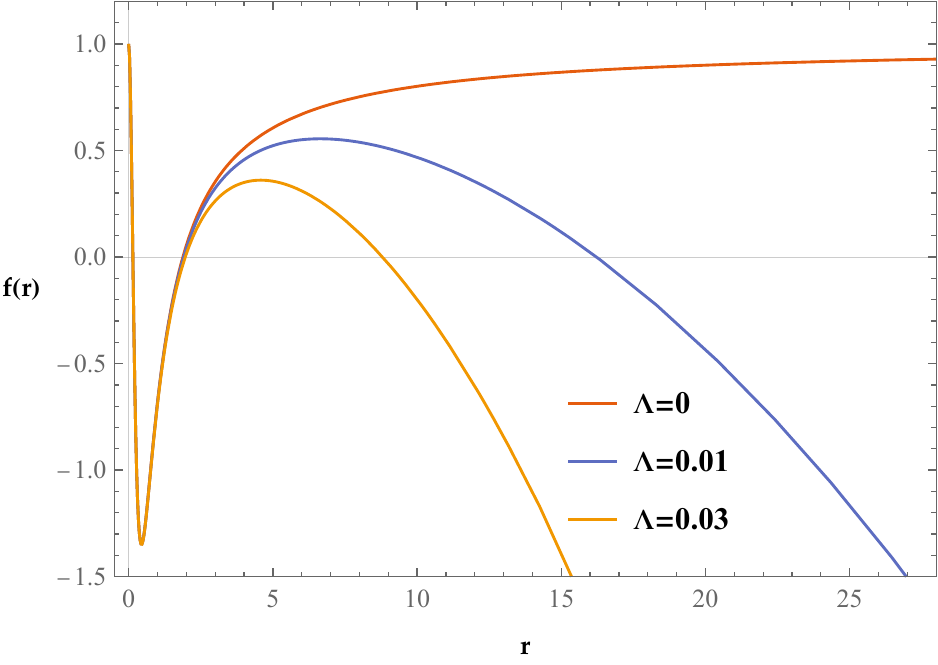}
    \caption{Profile of the metric function $f(r)$ for various $\Lambda$. We set $Q=0.3$. Increasing $\Lambda$ moves the cosmological horizon inward.}
    \label{dSf}
\end{figure}

\begin{figure}[!htbp]
    \centering
    \includegraphics[width=0.7\linewidth]{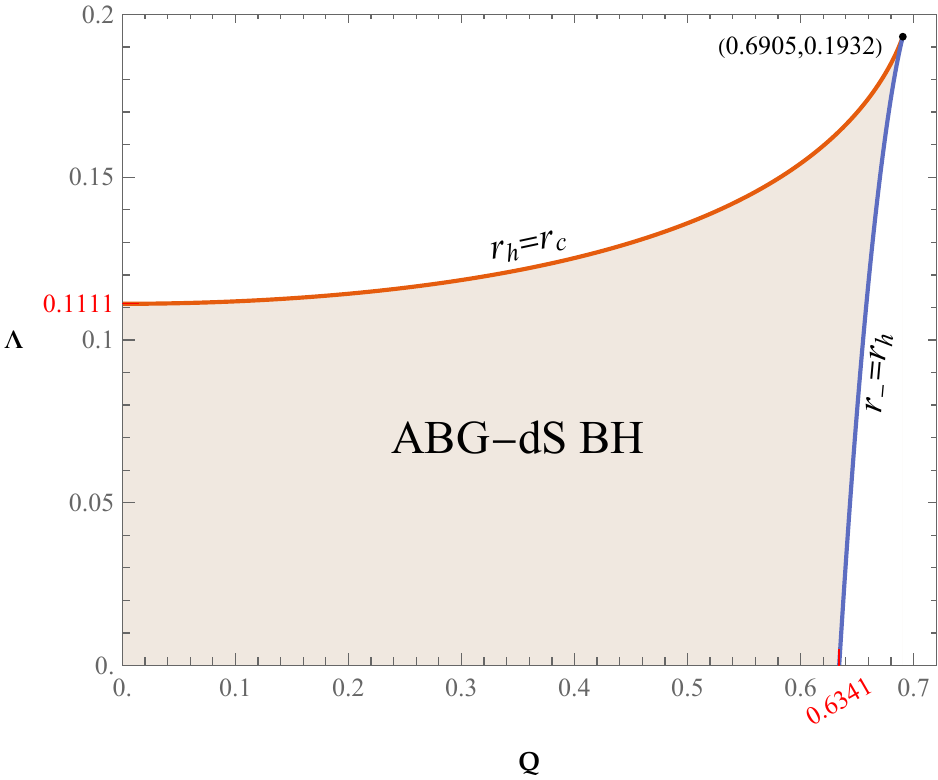}
    \caption{Allowed region in the $Q-\Lambda$ plane for the metric to describe a BH. The two critical curves $r_-=r_h$ and $r_h = r_c$ correspond to the two extreme limits. This domain fixes the parameter range used in the NER and efficiency analyses.}
    \label{LambdaQ}
\end{figure}

The metric of the Ay\'{o}n-Beato-Garc\'{i}a-de Sitter (ABG-dS) BH is \cite{Mo:2006tb}
\begin{align}
f(r) &= 1- \frac{2 M r^2}{(r^2 + Q^2)^{3/2}} + \frac{Q^2 r^2}{(r^2 + Q^2)^2} - \frac{\Lambda r^2}{3},\label{Metric}\\
A_t (r) &= - \frac{r^5}{2 Q} \left(\frac{3 M }{r^5} + \frac{2 Q^2}{(r^2 + Q^2)^3} - \frac{3 M}{(r^2 + Q^2)^{5/2}}\right), \label{ElectricPotential}
\end{align}
where $\Lambda$ is the cosmological constant. The typical profile of $f(r)$ is shown in Fig. \ref{dSf}. The electric potential $A_t$ does not depend on $\Lambda$, so its profile is the same as in the non-dS case (Fig. \ref{At-r}). From Fig. \ref{dSf}, one can see that there may be three horizons determined by $f(r)=0$: the inner horizon with $r=r_-$, the event horizon with $r=r_h$, and the cosmological horizon $r=r_c$. Moreover, the value of $r_c$ is significantly affected by $\Lambda$, with a larger $\Lambda$ leading to a smaller $r_c$. For the metric to describe a BH, we should have $r_- \leq r_h \leq r_c$, which thus limits the range of values of the parameters $\{Q, \Lambda\}$, as shown in Fig. \ref{LambdaQ}.

\begin{figure}[!htbp]
    \centering
    \includegraphics[width=0.7\linewidth]{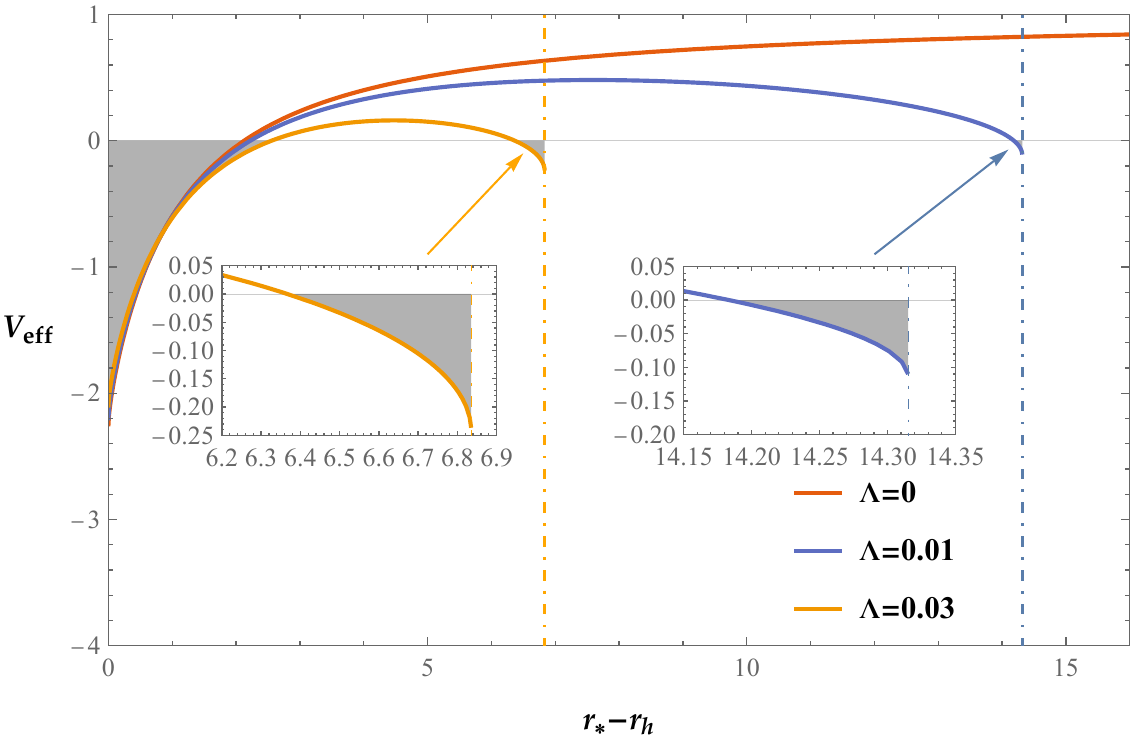}
    \caption{Profile of the effective potential $V_{\rm eff} (r)$ for various $\Lambda$. Vertical dashed lines denote the corresponding cosmological horizon $r=r_c$. We set $Q=0.3, \bar{q}=-5$ and $\ell=0$. A positive $\Lambda$ creates a second NER near $r_c$, which can join the event-horizon NER.}
    \label{VeffrLambda}
\end{figure}

\subsection{Negative-energy regions}

When the cosmological constant $\Lambda$ is incorporated, the effective potential and the corresponding NER undergo significant changes. Specifically, the NER emerges not only near the event horizon $r=r_h$ but also close to the cosmological horizon $r=r_c$, as shown in Fig. \ref{VeffrLambda}. This can also be seen from Eq.~(\ref{EffectivePotential}). When the cosmological constant $\Lambda$ is sufficiently large, the NER may extend throughout the entire region between $r_h$ and $r_c$, implying that energy extraction can occur at any location within the range $r_h<r<r_c$. Moreover, similar to the non-dS case, the existence of the NER requires the particle to carry a negative charge $q$.

\begin{figure}[!htbp]
	\centering
   \includegraphics[width=0.46\textwidth]{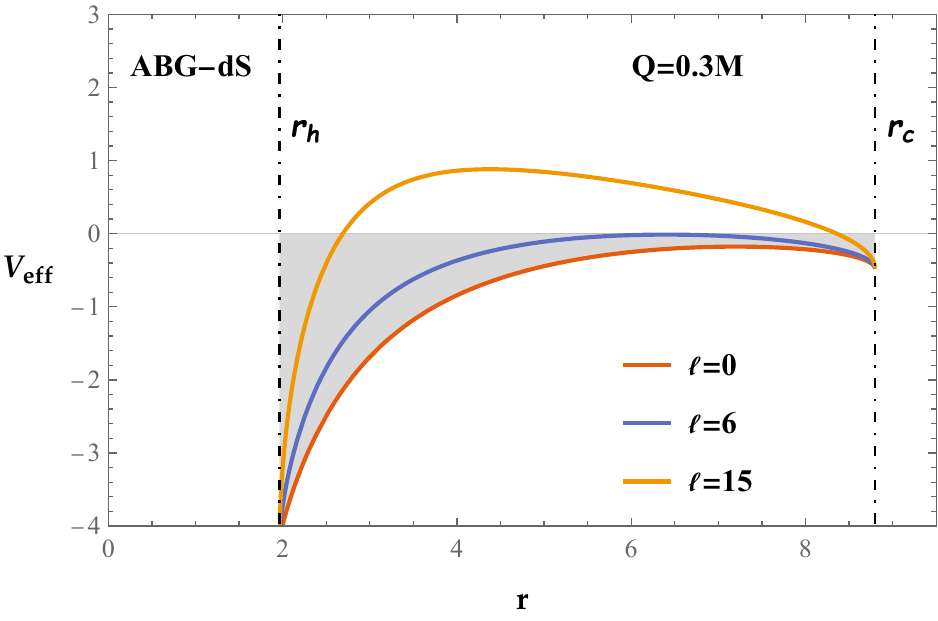}\quad 
    \includegraphics[width=0.46\textwidth]{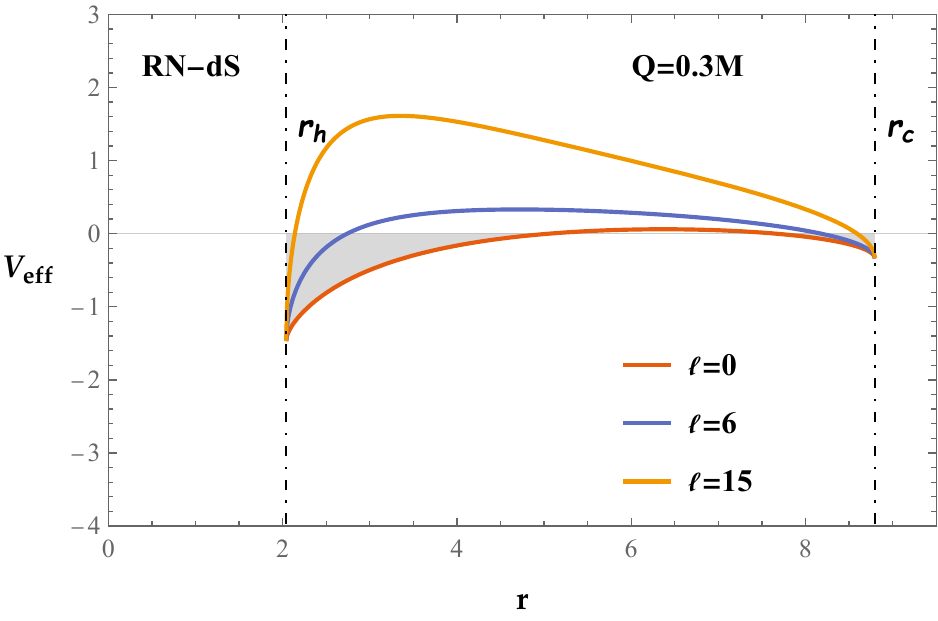}
    \caption{Effective potential $V_{\rm eff} (r)$ for the ABG-dS ({\em left panel}) and RN-dS ({\em right panel}) BHs for various $\ell$. The shaded gray regions indicate the NER. Vertical dashed lines denote the event horizon $r=r_h$ and the cosmological horizon $r=r_c$. We set $Q=0.3, \bar{q} = -10$, and $\Lambda =0.03.$ The comparison shows that the ABG-dS spacetime retains a larger NER than the RN-dS case.}
    \label{dsVeffNER}
\end{figure}

In Fig. \ref{dsVeffNER}, the profile of the effective potential $V_{\rm eff}(r)$ is plotted for various $\ell$. It can be observed that, for fixed $Q$, $\bar{q}$ and $\Lambda$, the NER expands as $\ell$ decreases and eventually encompasses the entire region between $r_h$ and $r_c$ when $\ell$ is below a certain value. In comparison with the non-dS case (Fig. \ref{VeffNER}), the presence of a positive cosmological constant results in a significantly larger NER. Furthermore, for fixed parameters, the NER in the ABG-dS case is always larger than that in the RN-dS case.

\subsection{Energy extraction efficiency}

\begin{figure}[!htbp]
    \centering
    \includegraphics[width=0.7\linewidth]{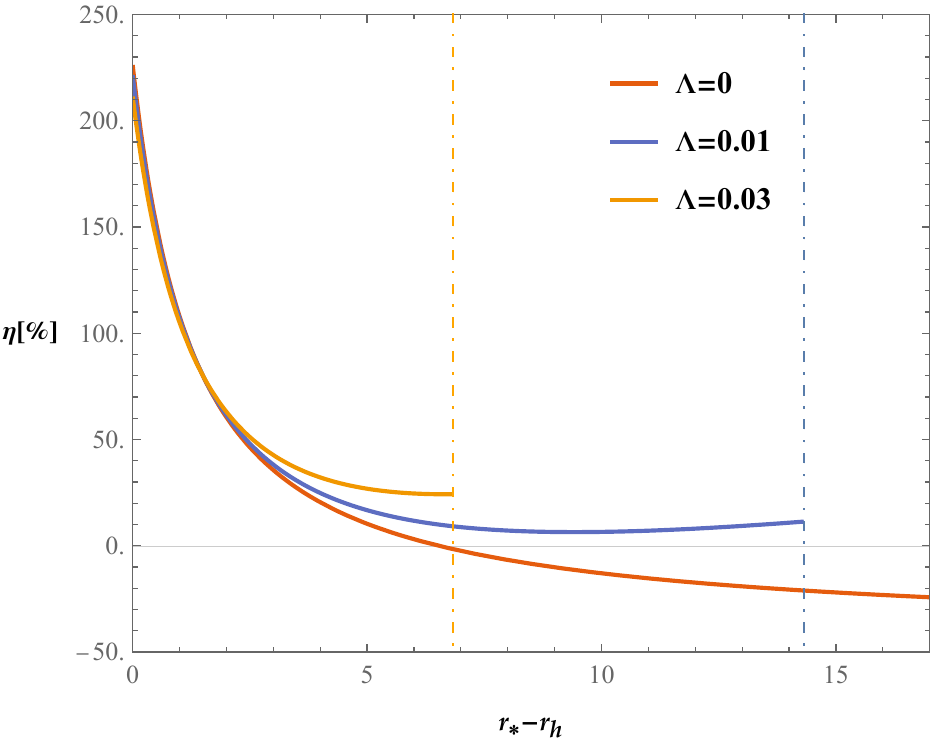}
    \caption{Energy extraction efficiency $\eta$ for the ABG-dS BH for various $\Lambda$. Vertical dashed lines denote the corresponding cosmological horizon. We set $Q=0.3$ and $\hat{\bar{q}}=-5$. A positive $\Lambda$ suppresses $\eta$ near $r_h$ but enhances it near $r_c$.}
    \label{etalambda}
\end{figure}

According to Eq.~(\ref{Efficiency-3}), in addition to the three parameters $\{Q, \hat{\bar{q}}, r_\ast\}$, the energy extraction efficiency $\eta$ now also depends on the cosmological constant $\Lambda$. In Fig.~\ref{etalambda}, $\eta$ is plotted as a function of the splitting point $r_\ast$ for various $\Lambda$. From the figure, it is evident that for fixed values of other parameters, when $\Lambda \neq 0$, $\eta$ is no longer a monotonically decreasing function of $r_\ast$ but rather a convex function. Similar to the non-dS case, $\eta$ attains its maximum value given in Eq.~(\ref{etaMax}) as $r_\ast \rightarrow r_h$. Moreover, near the event horizon, a larger $\Lambda$ results in lower efficiency. Conversely, near the cosmological horizon, efficiency increases as $\Lambda$ increases.

\begin{figure}[!htbp]
    \centering
    \includegraphics[width=0.7\linewidth]{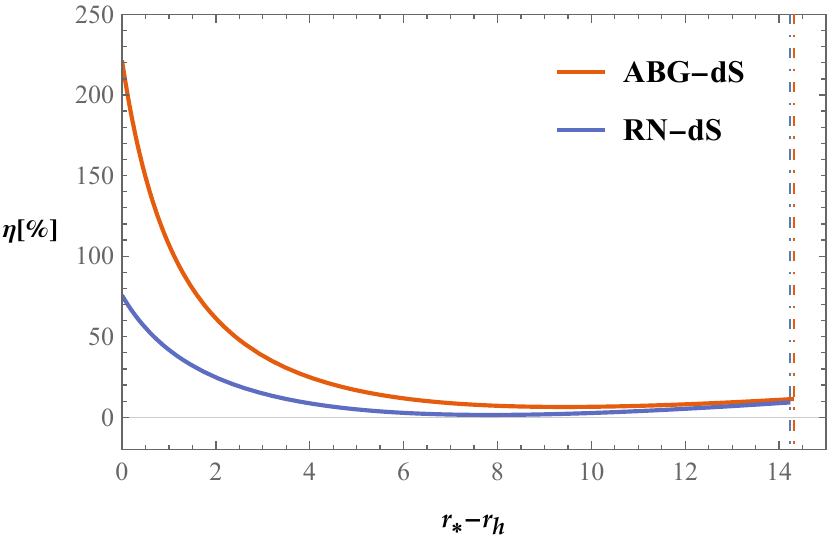}
    \caption{Energy extraction efficiency $\eta$ as a function of the splitting point $r_\ast$ for the ABG-dS and RN-dS cases, with $Q=0.3M$, $\hat{\bar{q}}=-5$ and $\Lambda=0.01$. Vertical dashed lines denote the corresponding cosmological horizon $r=r_c$. The ABG-dS curve remains above the RN-dS curve, indicating stronger extraction at the same splitting radius.}
    \label{dSetar}
\end{figure}

In Fig. \ref{dSetar}, we compare the energy extraction efficiencies of the ABG-dS case and RN-dS case. It is evident that, for fixed parameters, $\eta$ for the ABG-dS case is always higher than the RN-dS case, and the difference becomes larger as $r_\ast$ decreases. 

\begin{figure}[!htbp]
    \centering
    \includegraphics[width=0.7\linewidth]{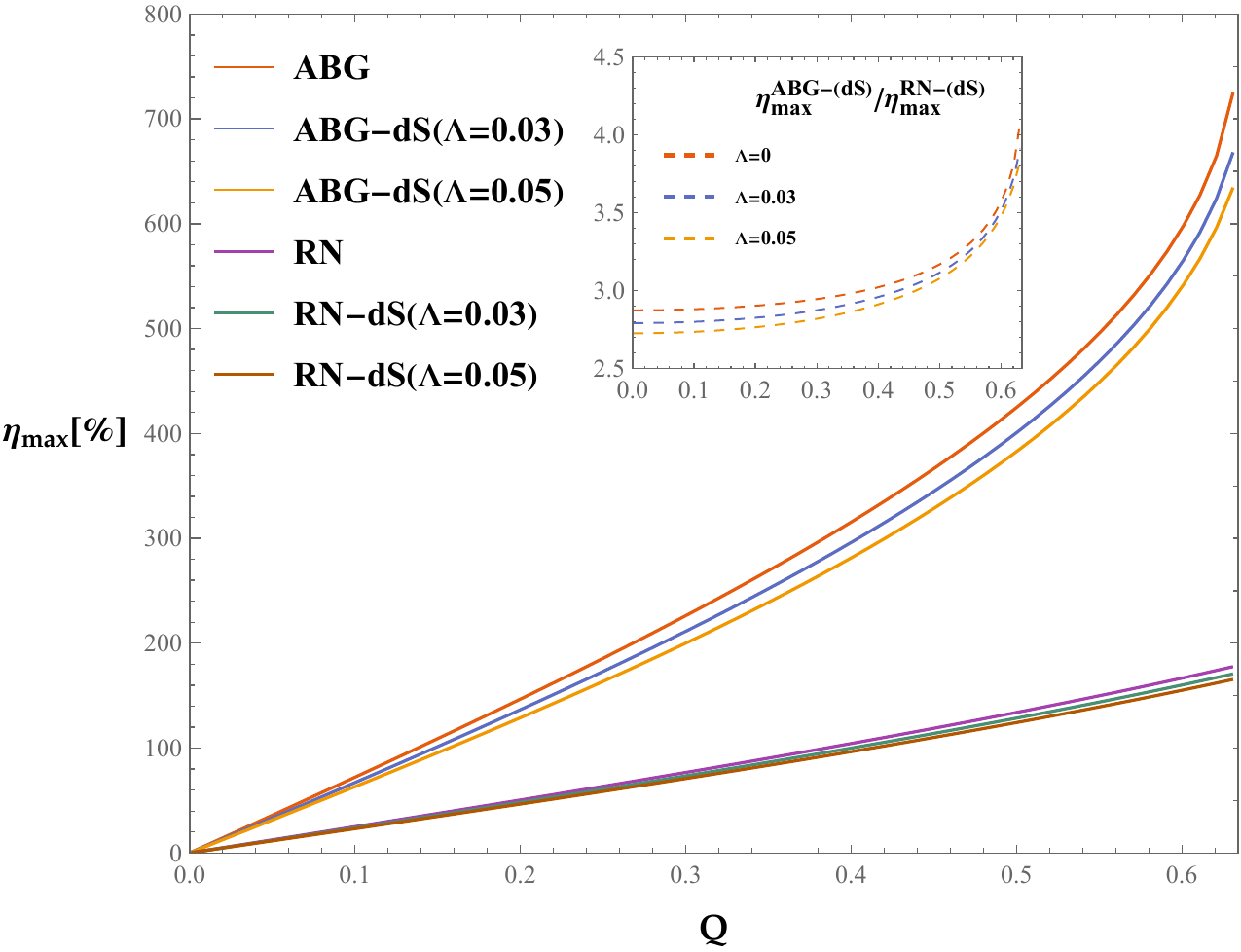}
    \caption{The maximum energy extraction efficiency $\eta_{\rm max}$ as a function of the BH charge $Q$, with the splitting point $r_\ast = r_h$. We set $\hat{\bar{q}}=-5$. The trend shows that increasing charge enhances the maximum efficiency, while a positive $\Lambda$ lowers it.}
    \label{dSetaQ}
\end{figure}

In Fig. \ref{dSetaQ}, the influence of the black hole charge $Q$ on the maximum energy extraction efficiency $\eta_{\rm max}$ is illustrated for various cases. It can be observed that $\eta$ is a monotonically increasing function of $Q$. For fixed parameters, $\eta_{\rm max}$ in the ABG(-dS) case is significantly higher than that in the RN(-dS) case. Furthermore, a larger $\Lambda$ results in a lower $\eta_{\rm max}$. Consistent with our discussion in the previous section, for the very small $Q$ and $\Lambda$ values typical in realistic astrophysical contexts, the ratio $\eta_{\rm max}^{\rm ABG-(dS)} / \eta_{\rm max}^{\rm RN-(dS)}$ is approximately $23/8$.

\section{Summary and discussions}

In this work, we have systematically investigated the electric Penrose process in a class of regular black holes, specifically focusing on the ABG and ABG-dS black hole solutions. We have derived the general formalisms governing the effective potential and energy extraction efficiency, and conducted a comprehensive analysis of how various physical parameters influence the negative-energy region and the efficiency of energy extraction. Furthermore, we have performed detailed comparisons with the RN and RN-dS black hole cases to elucidate the fundamental differences between regular and singular black hole spacetimes. Our principal findings are summarized as follows:

\begin{itemize}
    \item For fixed physical parameters, the negative-energy region in ABG(-dS) black holes is substantially larger than that in RN(-dS) black holes, as clearly demonstrated in Figs. \ref{VeffNER} and \ref{dsVeffNER}. This expanded region enables the electric Penrose process to occur at significantly greater distances from the horizons, enhancing the astrophysical relevance of this mechanism.

    \item For identical parameter values, the energy extraction efficiency $\eta$ in ABG(-dS) black holes exceeds that in RN(-dS) black holes, particularly in the vicinity of the event horizon, as illustrated in Figs. \ref{etarABGRN} and \ref{dSetar}. This trend is also observed for the maximum extraction efficiency $\eta_{\rm max}$, with the discrepancy between the two cases becoming increasingly pronounced as the black hole charge $Q$ increases (Figs. \ref{etaQ} and \ref{dSetaQ}).

    \item Even for the extremely small values of $Q$ and $\Lambda$ characteristic of realistic astrophysical scenarios \footnote{In realistic astrophysical environments, observational constraints limit the charge to the range $33.4 \frac{M}{M_\odot} {\rm C} \lesssim Q \lesssim 3.34\times 10^{8} \frac{M}{M_\odot} {\rm C}$ ~\cite{Zajacek:2018ycb, Zajacek:2019kla}. In our dimensionless units (with $M=1$), this corresponds to $1.94 \times 10^{-19} \lesssim Q \lesssim 1.94 \times 10^{-12} $, which is many orders of magnitude smaller than the theoretical maximum. Additionally, current cosmological observations indicate $\Lambda \sim 1.1 \times 10^{-52} \ m^{-2}$ \cite{Planck:2018vyg, DES:2018rjw, Knobles:2025xeo, DES:2024ywx}, translating to $\Lambda \sim 2.4 \times 10^{-46} \left(\frac{M}{M_\odot}\right)^2$ in our units - an exceedingly small value for typical astrophysical black holes.}, \rev{the leading-order ABG(-dS)/RN(-dS) efficiency ratio remains finite.} Specifically, we obtain the relation:
\begin{align}
    \eta_{\rm max}^{\rm ABG-(dS)} / \eta_{\rm max}^{\rm RN-(dS)} \approx 23/8, \qquad {\rm for \ small} \ Q \ {\rm and } \ \Lambda.
\end{align}
\rev{This ratio characterizes the idealized single-particle efficiency and should not be interpreted as a statistical significance or a prediction of an observable luminosity.}

    \item  When the cosmological constant $\Lambda$ is included, the electric Penrose process exhibits novel features: it can occur not only near the event horizon but also in the vicinity of the cosmological horizon. Moreover, in this extended scenario, $\eta$ is no longer a monotonically decreasing function of the splitting point $r_\ast$, but rather displays a convex functional form (Fig. \ref{etalambda}). Specifically, near the event horizon, a larger $\Lambda$ results in reduced efficiency $\eta$, while conversely, near the cosmological horizon, a larger $\Lambda$ leads to enhanced $\eta$.
\end{itemize}

\rev{The assumption $q_1=0$ selects a particular ionization or decay channel in which a neutral parent splits into oppositely charged fragments; it does not require incident particles in general to be neutral. Likewise, $e_1\simeq1$ specifies a marginally bound parent rather than a necessary condition for the process. For the same $E_1$ and $q_2$, the horizon-limit ABG--RN ratio is controlled by the corresponding horizon potentials and therefore does not require $q_1=0$ or $e_1=1$, although finite-radius efficiencies remain channel dependent.}

\rev{The quoted efficiencies are single-event kinematic results, and a large $\eta$ does not by itself imply a large absolute energy transfer.}

\rev{These conclusions rely on the test-particle approximation and neglect backreaction, self-force effects, radiation losses, and changes in the background electromagnetic field. Provided these corrections remain perturbative, they may reduce the numerical efficiencies, especially for particles with very large charge-to-mass ratios, but are not expected to overturn the qualitative ABG(-dS)--RN(-dS) comparison, which is determined mainly by the background metric and electric potential.}

In summary, our comprehensive analysis suggests that ABG black holes possess superior energy extraction characteristics compared to RN black holes in both asymptotically flat and de Sitter spacetimes. These findings significantly advance our understanding of the physical properties of regular black holes and provide crucial theoretical insights into the role of nonlinear electrodynamics corrections in astrophysical energy extraction processes. Future research directions include extending this analysis to other classes of regular black hole solutions, incorporating rotational effects, and examining the influence of magnetic fields on the electric Penrose process.

\begin{acknowledgments}

	This work is supported by the National Natural Science Foundation of China (NNSFC) under Grant No 12075207. \rev{We would like to thank the anonymous referees for their valuable and constructive suggestions, which have significantly improved the quality of this manuscript.}

\end{acknowledgments}
\vspace{0.5cm}


\bibliographystyle{utphys}
\bibliography{ref}

\end{document}